\documentclass[prl,superscriptaddress,twocolumn,showpacs,preprintnumbers,amsmath,amssymb]{revtex4}
\usepackage{graphicx}% Include figure files
\usepackage{dcolumn}% Align table columns on decimal point
\usepackage{bm}% bold math
\begin{document}

\preprint{APS/123-QED}

\title{Ferromagnetically coupled Shastry-Sutherland quantum spin singlets in (CuCl)LaNb$_2$O$_7$}% Force line breaks with \\

\author{C. Tassel}
\affiliation{Department of Energy and Hydrocarbon Chemistry, Graduate School of Engineering, Kyoto University, Kyoto 615-8510, Japan}
\author{J. Kang}
\author{C. Lee}
\affiliation{Department of Chemistry, North Carolina State University, Raleigh, North Carolina 27695-8204, USA}
\author{O. Hernandez}
\affiliation{University of Rennes 1, Sciences Chimiques de Rennes UMR CNRS 6226, Campus de Beaulieu Batiment 10B, 35042 Rennes, France}
\author{Y. Qiu}
\affiliation{NIST Center for Neutron Research, National Institute of Standards and Technology, Gaithersburg, MD 20899, USA}
\author{W. Paulus}
\affiliation{University of Rennes 1, Sciences Chimiques de Rennes UMR CNRS 6226, Campus de Beaulieu Batiment 10B, 35042 Rennes, France}
\author{E. Collet}
%\affiliation{University of Rennes 1, Sciences Chimiques de Rennes UMR CNRS 6226, Campus de Beaulieu Batiment 10B, 35042 Rennes, France}
\affiliation{University of Rennes 1, Institut de Physique de Rennes UMR CNRS 6251, Campus de Beaulieu Batiment 11A, 35042 Rennes, France}
\author{B. Lake}
\email{bella.lake@helmholtz-berlin.de}
\affiliation{Helmholtz-Zentrum Berlin, Berlin 14109, Germany}
\affiliation{Institut fur Festkorperphysik, Technische Universitat Berlin, Hardenbergstrabe 36, 10623 Berlin, Germany}
\author{T. Guidi}
\affiliation{Helmholtz-Zentrum Berlin, Berlin 14109, Germany}
\affiliation{ISIS Facility, Rutherford Appleton Laboratory, Chilton, Didcot, Oxon OX11 0QX, United Kingdom}
\author{M.-H. Whangbo}
\affiliation{Department of Chemistry, North Carolina State University, Raleigh, North Carolina 27695-8204, USA}
\author{C. Ritter}
\affiliation{Institute Laue Langevin, BP 156, 38042, Grenoble, France}
\author{H. Kageyama}
\email{kage@scl.kyoto-u.ac.jp}
\affiliation{Department of Energy and Hydrocarbon Chemistry, Graduate School of Engineering, Kyoto University, Kyoto 615-8510, Japan}
\author{S.-H. Lee}
\email{shlee@virginia.edu}
\affiliation{Department of Physics, University of Virginia, Charlottesville, VA 22904-4714, USA}
\date{\today}% It is always \today, today,

\begin{abstract}
 Using inelastic neutron scattering, x-ray, neutron diffraction, and the first-principle calculation techniques, we show that the crystal structure of the two-dimensional quantum spin system (CuCl)LaNb$_2$O$_7$ is orthorhombic with $Pbam$ symmetry in which CuCl$_4$O$_2$ octahedra are tilted from their high symmetry positions and the Cu$^{2+} (s = 1/2)$ ions form a distorted square lattice. The dominant magnetic interactions are the fourth nearest neighbor antiferromagnetic interactions with a Cu-Cl--Cl-Cu exchange path, which lead to the formation of spin singlets. The two strongest interactions between the singlets are ferromagnetic, which makes (CuCl)LaNb$_2$O$_7$ the first system of ferromagnetically coupled Shastry-Sutherland quantum spin singlets.
\end{abstract}

\pacs{Valid PACS appear here}% PACS, the Physics and Astronomy
\maketitle

Quantum magnets continue to attract interest due to the wide range of exotic ground states and excitations that they display.\cite{sachdev} Such magnets usually have magnetic ions with low spin values (typically spin-1/2 or 1) which are coupled by low-dimensional and/or frustrated magnetic interactions.
A quantum magnet that has attracted a lot of theoretical interest is the square lattice of spin-1/2 ions where the first ($J_1$) and second $(J_2)$ neighbor interactions which couple the magnetic ions along the sides and diagonals of the square lattice, respectively, compete with each other.\cite{sachdev2, rosner} However in many cases the predictions have not been verified experimentally due to the lack of physical realizations of this system. Recently (CuCl)LaNb$_2$O$_7$ was proposed as an example of frustrated square lattice. Here the magnetic Cu$^{2+}$ ions which possess spin moment S=1/2 are octahedrally co-ordinated and form CuCl planes well separated by LaNb$_{2}$O$_{7}$ layers. The crystal symmetry was originally found to be tetragonal with space group $P4/mmm$ and one copper ion per unit cell $(a_t = b_t = 3.88$ \AA) forming an ideal undistorted square lattice of S = 1/2 ions with uniform $J_1$ and $J_2$.\cite{TAK99}
Theory suggests that (CuCl)LaNb$_2$O$_7$ would have long-range magnetic order which is ferromagnetic or antiferromagnetic depending on the ratio of the two exchange constants.\cite{schmidt} Gapped excitations are not expected except for a very limited region of the phase diagram when the ratio of interactions along the diagonal to those along the square are $\approx$ 0.5\cite{sindzingre} while a gapless spin disordered phase is predicted when this ratio is $\approx$ -0.5.\cite{shannon} However the experimental data appears to contradict the theory, heat capacity and magnetization measurements suggested that the ground state is in fact a singlet and that the excitations are gapped.\cite{kageyama2}  Neutron scattering also revealed the absence of long range magnetic order and showed that the excitations are gapped and centered at 2.3 meV.\cite{kageyama} Such behavior is typical of a dimerised magnet where a dominant antiferromagnetic exchange interaction couples the spins into dimer pairs with no net moment and a finite energy is required to break up the dimers and create a triplet ($S = 1$) excitation. However such dimerization is incompatible with the high crystal symmetry where $J_1$ and $J_2$ are uniform and the spins cannot be paired. 

This paper describes a reinvestigation of the crystal structure and shows that the space group is in fact orthorhombic ($Pbam$) with a unit cell that is doubled along both directions within the $ab$-plane compared to the original tetragonal symmetry, thus allowing for dimerisation. The Cu-Cl bonds are considerably distorted and band structure calculations reveal highly spatially anisotropic exchange interactions and predict dimerisation between fourth neighbour ions, in agreement with our inelastic neutron scattering data. The resulting magnetic structure is best described as antiferromagnetic dimers which are coupled together by frustrated ferromagnetic interactions in a Shastry-Sutherland type arrangement.

%-----------------------------------------
%	Introduction
%-----------------------------------------

% Sample Preparation 
A 15 g powder sample of (CuCl)LaNb$_2$O$_7$ was synthesized by an ion-exchange reaction using a powder mixture of  RbLaNb$_2$O$_7$ and CuCl$_2$ with a molar ratio of 1:2 heated at 335 $^o$C for one week, as described previously \cite{TAK99}. 
Concerning single-crystal samples, precursor CsLaNb$_2$O$_7$ single crystals (see Ref. \cite{NK96}) with typical size of 0.5 $\times$ 0.5 $\times$ 0.1 mm$^3$ were embedded in a molar excess of CuCl$_2$ powder (99.99\%) sealed in an evacuated tube and reacted at 340 $^o$C for one week. The mixture was then washed with distilled water to remove the byproducts and unreacted CuCl$_2$. The square, plate-like dark green crystals of (CuCl)LaNb$_2$O$_7$ were then dried overnight at 120 $^o$C. The Energy Dispersive Spectroscopy (EDS) study revealed the absence of the Cs atoms, indicating that the ion exchange reaction occurs completely even for the single crystal.

% Structural Characterization. 
A CCD and a 4-circle diffractometers  were used for the
single-crystal data collection at 293 and 14 K, respectively. Diffraction patterns of the powder
were measured at room and low temperature using synchrotron light with $\lambda = 0.77749$ \AA~at BL02B2  beamline at SPring-8 as well as neutron beam with $\lambda = 1.9085$ \AA~at D1A at ILL.  Powder inelastic neutron scattering measurements were done with $\lambda =$ 3.8 \AA~at the Disk-Chopper-Spectrometer (DCS) at NCNR, NIST under an external magnetic field.  The crystal structure was solved by direct methods against single crystal data and further refined by taking into account the twinning due to the pseudo-symmetry. The structural parameters were optimized by a Reitveld refinement against neutron diffraction data, more sensitive for O and Cl atoms. All the refinements were carried out by the means of the JANA2000 refinement program \cite{jana}.

 %-----------------------------------------
%	Figure 1.
%-----------------------------------------
\begin{figure}
\includegraphics[width=0.48\textwidth]{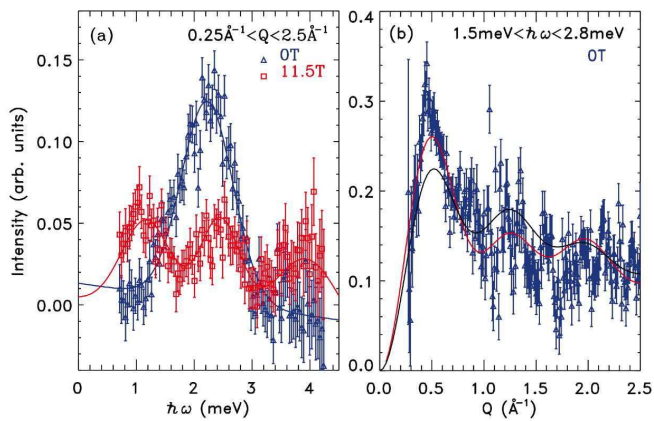}
\centering
\caption{Inelastic neutron scattering data obtained at 70 mK from the powder sample. (a) Energy dependence of the triplet excitations for $B=0$ T (blue) and $B=11.5$ T (red). (b) $Q$-dependence for $B=0$ T. Lines are described in the text.
\label{fig1}}
\end{figure}

Fig. 1 (a) shows the inelastic neutron scattering data as a function of energy transfer, $\hbar\omega$. In the absence of an external magnetic field, $B=0$ T, a single peak is observed centered at $\hbar\omega = 2.22(1)$ meV in agreement with previous work.\cite{kageyama} An external magnetic field of $B = 11.5$ T splits this peak into three peaks at $\hbar\omega =$ 1.19(1), 2.45(3), and 3.95(11) meV. The energies are consistent with the Zeeman splitting expected for a magnon excitation with spin S = 1 ($g\mu_B S_z B$) thus confirming that the excitations are due to a transition from a singlet ground state to a triplet excited state and implying that the spins in (CuCl)LaNb$_2$O$_7$ are paired into dimers where the dimer exchange constant is $\approx 2.22$ meV. In zero field the excitation extends in energy from $\sim$ 1.2 meV to $\sim$ 3.0 meV and is thus much broader than the resolution (0.16 meV) implying that it is dispersive due to interactions between dimers. Indeed the lower edge of the excitation is consistent with the critical field of $B_{c1}$ = 10.3 T from magnetization measurements, for condensation of magnons into the ground state.\cite{kageyama2}

The nature of the spin dimers in (CuCl)LaNb$_{2}$O${_7}$ can be obtained from the wavevector-dependence of the triplet excitations, $I(Q)$. As shown in Fig. 1 (b), the $Q$-dependence exhibits a prominent peak centered at $Q_{c} \approx0.5$ \AA$^{-1}$. If we assume the dimers are non-interacting, the $Q$-dependence of their excitation goes as, $1-\frac{{\rm sin} Qr}{Qr}$ where $r$ is the intradimer distance. The best fit of the data to the isolated spin dimer model was obtained with $r = 8.4(2)$ \AA~(black line in Fig. 1 (b)) in agreement with previous work.\cite{kageyama} The model reproduces the observed peak position of $Q_c \sim$ 0.5 \AA$^{-1}$ well. This reveals that the dominant magnetic interactions are between the Cu$^{2+}$ ions that are separated much further than the nearest neighbor distance of 3.88 \AA, a result that is inconsistent with the currently accepted structure of (CuCl)LaNb$_2$O$_7$. 

 %-----------------------------------------
%	Figure 2.
%-----------------------------------------
\begin{figure}
\includegraphics[width=0.48\textwidth]{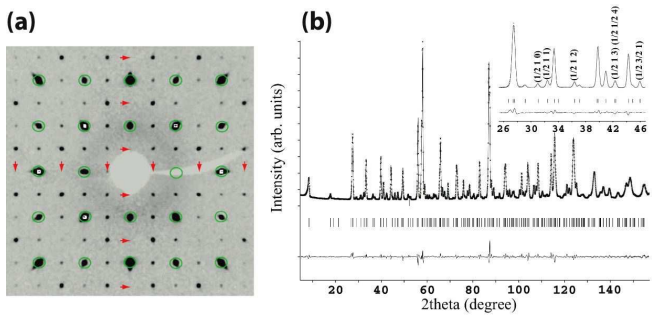}
\centering
\caption{Single crystal laboratory X-ray and powder neutron diffraction data of (CuCl)LaNb$_2$O$_7$. (a) the X-ray CCD image of the $ab$-plane at 14 K. The single crystal was twinned, thus $a$- and $b$-axes are interchangable. The spots with green circles are the allowed integer Bragg reflections for the original tetragonal structure, the other spots are the half-integer reflections. 
(b) Neutron diffraction pattern at 2 K. Dots are the data, solid line is the fit to the orthorhombic structure, and the line on the bottom panel is the difference between the data and the model. The indices of the reflections in the inset are in the tetragonal notation.
\label{fig2}}
\end{figure}

%-----------------------------------------
%	Table I.
%-----------------------------------------
\begin{table}
  \caption{Structural parameters obtained from Rietveld refinement of neutron diffraction at 2 K. 
 The goodness of the fit is indicated by the conventional structural R$_{Bragg}$, $R_{wp}$, $R_{p}$, and $\chi^2$.
 %The goodness of the fit can be measured by the following quantities:
 %$R_{wp} = \sqrt{ \sum w_i (y_{io}-y_{ic})^2 / \sum w_i y_{io}^2}, R_p = \sum |y_{io}-y_{ic}| /\sum y_{io}, \chi^2 = \sum w_i (y_{io}-y_{ic})^2 / (N-P)$, where $y_{io}$ and $y_{ic}$ are the observed and calculated intensities, $w_i$ is the weighting factor that is the inverse of the error, $N$ is the total number of the data, and $P$ is the number of the fitted parameters.
  \label{table1}}
  \begin{ruledtabular}
%    \newcolumntype{.}{D{.}{.}{-1}}
    \begin{tabular}{ccccc}
      \multicolumn{5}{l}{$Pbam$, $a$ = 7.7556(5) {\AA}, $b$ = 7.7507(5) {\AA}, $c$ = 11.7142(4) {\AA}} \\
      \multicolumn{5}{l}{$\chi^2 = 3.38 $, R$_{Bragg}$ = 1.64\%, R$_p$ = 3.39 \%, R$_{wp}$ = 4.62 \%} \\ \hline
      Atom($W$) & $x/a$ & $y/b$ & $z/c$ & $U_{\textrm{iso}}$({\AA}$^2$) \\ 
     Cu (4$h$) & 0.502(6) & 0.7662(10) & 0.5 & 0.0043(13) \\
     Cl (4$h$) & 0.7367(16) & 0.4286(7) & 0.5 & 0.006(2) \\
     La (4$g$) & 0.7412(11) & 0.5001(10) & 0 & 0.0024(12) \\
     Nb (8$i$) & 0.500(3) & 0.7529(9) & 0.1908(3) & 0.0040(9) \\
     O1 (4$f$) & 0.5 & 0 & 0.1397(13) & 0.002(3) \\
     O2 (4$e$) & 0.5 & 0.5 & 0.1751(11) & 0.008(3) \\
     O3 (8$i$) & 0.752(4) & 0.7509(11) & 0.1545(10) & 0.006(2) \\
     O4 (4$g$) & 0.500(2) & 0.7170(14) & 0 & 0.007(2) \\
     O5 (8$i$) & 0.501(4) & 0.7731(11) & 0.3410(5) & 0.0089(16) \\
     \end{tabular}
  \end{ruledtabular}
\end{table}

 %-----------------------------------------
%	Figure 3
%-----------------------------------------
\begin{figure}
\includegraphics[width=0.48\textwidth]{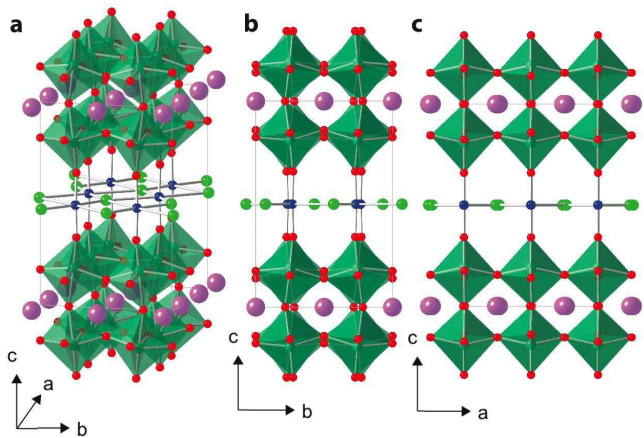}
\centering
\caption{Crystal struture of (CuCl)LaNb$_2$O$_7$. (a) Three dimensional view, (b) the $bc$-projection, and (c) the $ac$-projection. Purple, red, blue and green spheres represent La, O, Cu and Cl atoms, respectively. The dark green polygons represent NbO$_6$ octahedra.  The grey lines show the unit cell.
\label{fig1}}
\end{figure}

% Results: Structures  
To understand this unusual $Q$-dependence, we have reinvestigated the crystal structure in great detail.
Fig. 2 (a) shows the CCD image of the diffraction data taken for the $(hk0)$ plane at 14 K. Note that the crystal is twinned and thus $h$ and $k$ are interchangeable. In addition to the allowed integer peaks for the $P4/mmm$ tetragonal structure, half-integer superlattice spots are also found, indicating that the real unit cell is  $2a_t \times 2b_t \times c$ compared to the tetragonal unit cell. Similar superreflections are also observed in the powder synchrotron x-ray and neutron diffraction (see Fig. 2 (b)). Furthermore, the CCD image shows that the superlattice peaks (odd $h$,0,0) and (0,odd $k$,0) are absent in the orthorhombic notation (red arrows). This extinction condition tells us that the crystal structure is orthorhombic with either $Pbam$ or $Pba2$ space group. The best fit to the x-ray and neutron diffraction data was obtained with the $Pbam$ structure. Table I lists the optimal structural parameters from the neutron diffraction data that are more sensitive to the Cl and O atoms than the x-ray data.  The validity of the structural analysis is further supported by the bond valence sum calculation for copper yielding +2.02. A comparison of diffraction patterns at different temperatures revealed that no structural phase transition occurs below 293 K. 

 %-----------------------------------------
%	Figure 4
%-----------------------------------------

%\begin{figure}
%\includegraphics[width=0.48\textwidth]{NbO6_Cu.eps}
%\centering
%\caption{Local distortion of the NbO$_6$ and CuCl$_4$O$_2$ octahedra, and magnetic orbital. (a) NbO$_6$ octahedron. (b) CuO$_2$Cl$_4$ octahedron. $x$, $y$, $z$ represent the local axes. (c) Magnetic orbital of the axially-elongated CuO$_2$Cl$_4$ octahedron. In (b) and (c), the local z- and x-axes of the octahedron are taken along the Cu-O and the short Cu-Cl bonds, respectively. 
%} 
%\label{fig1}}
%\end{figure}

  %-----------------------------------------
%	Figure 5
%-----------------------------------------
\begin{figure}
\includegraphics[width=0.48\textwidth]{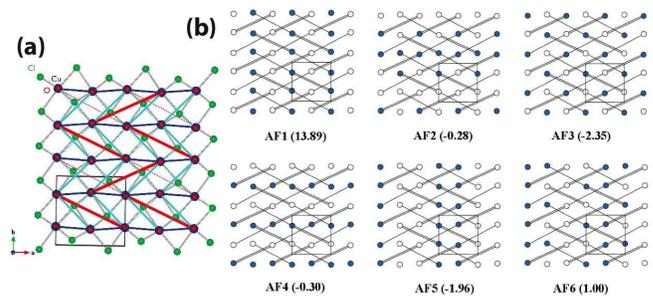}
\centering
\caption{(a) Magnetic exchange interactions in the $ab$-plane of (CuCl)LaNb$_2$O$_7$. 
%A distorted square lattice of the Cu$^{2+}$ ions with quantum spin $s = 1/2$ is formed in the $ab$-plane formed by Cu and Cl atoms. 
The blue, green, and red spheres represent the Cu, Cl, and O ions, respectively. The O atoms are located on the top and bottom of the Cu atoms. The lines connecting Cu atoms represent exchange bonds: $J_{1a}$ (blue), $J_{2a}$ (solid cyan), $J_{2b}$ (dotted cyan), $J_4$ (solid red), and $J_4^{'}$ (dotted red line). 
%The thickness of the lines reflects the strength of the bonds. 
See Table II. The negligible nearest neighbor interaction along the $b$-axis, $J_{1b}$, is not plotted for clarity.
(b) Six ordered antiferromagnetic spin states used to extract the spin exchange parameters by GGA+U calculations. Only the Cu$^{2+}$ ions are shown, and the open and filled circles represent the up- and down-spins, respectively. The relative energies of the ordered spin states are given in the parentheses.} 
%\label{fig1}}
\end{figure}

%-----------------------------------------
%	Table II.
%-----------------------------------------
\begin{table}
  \caption{%Relative coupling constants and corresponding exchange paths for up to the 4th nearest neighbor interactions in (CuCl)LaNb$_2$O$_7$. $J_4$ is antiferromagnetic and its strength, determined from the inelastic neutron scattering (Fig. 1 (a)) is $J_4 \sim 2.22$ meV. The negative ratios of $J$s/$J_4$ means those $J$s are ferromagnetic.
 Exchange paths for up to the 4th nearest neighbor interactions in (CuCl)LaNb$_2$O$_7$ and the corresponding coupling constants relative to $J_{4}$. $J_{4}$ is antiferromagnetic and of strength $\sim 2.22$ meV. Negative ratio $J_s$/$J_4$ means that $J_s$ is ferromagnetic. Short and long lines between atoms represent different bond lengths without proportion.
   \label{table2}}
  \begin{ruledtabular}
    \begin{tabular}{ccccc}
     $J$s & path & $d$ (\AA) & angle $(^o)$ & $J$s/$J_4$ \\ \hline
     $J_{1a}$ & Cu-Cl-Cu & 3.88548 & 108.9, 75.8 & -0.39  \\
     $J_{1b}$ & Cu-Cl--Cu & 3.88548 & 80.9  & -0.04 \\
     $J_{2a}$ & Cu-Cl--Cu &  5.46148 & 156.7 & -0.38\\
     $J_{2b}$ & Cu-Cl--Cu & 5.5053 &  170.2 & -0.14 \\
     $J_{4}$ & Cu-Cl--Cl-Cu & 8.81262 & 164.9 & 1 \\
     $J_{4}^{'}$ & Cu-Cl--Cl-Cu & 8.53250 & 150.0 & 0.18 \\
      \end{tabular}
  \end{ruledtabular}
\end{table}

The refined orthorhombic crystal structure differs markedly from the tetragonal structure previously reported. While the NbO$_6$ octahedra align with the axes in the tetragonal structure they are strongly tilted in the orthorhombic structure particularly around the $a$-axis in a staggered manner. 
%(counter-clock-wise and clock-wise alternating) along both the $b$- and the $a$-axis. 
Furthermore, the Nb ions are shifted along the $c$-axis from the center of the octahedron. The tilting pattern of the NbO$_6$ strongly influences the positions of both the copper and chlorine atoms. In particular, the chlorine atoms move significantly along the $b$-direction and slightly along the $a$-direction from their tetragonal position, probably to reduce Coulomb repulsion between chlorine and apical oxygen atoms. The Cu ions occupy the $4h$ sites. Along the $a$-direction, they are mostly transversally displaced, i.e., along the $b$-axis, yielding the distance between the nearest neighbor Cu-Cu ions to be 3.626 \AA~and 4.129 \AA~along the $b$-axis and 3.885 \AA~along the $a$-axis. The copper ion is coordinated octahedrally by two oxygen ligands with a distance of 1.865 \AA~as well as four chlorine ligands with two shorter bonds (2.386 \AA, 2.389 \AA) and two longer bonds (3.136 \AA, 3.188 \AA). When the local $z$- and $x$-axes for each Cu$^{2+}$ ion are taken along the Cu-O and the short Cu-Cl bonds, respectively, the overall symmetry of the magnetic orbital including the ligand $p$-orbitals has the $z^2-x^2$ character. As a result, the spin exchange interactions in the CuCl layer become highly anisotropic. 
%The Cu-Cl-Cu angles for the diagonal coppers are 157.12$^o$ and 170.42$^o$.

%Noting that the strong exchange interactions between Cu$^{2+}$ ions should involve exchange paths containing the Cu-Cl bonds, 
As for the spin exchanges of (CuCl)LaNb$_2$O$_7$, we consider Cu-Cl-Cu and Cu-Cl--Cl-Cu superexchanges. In general, one expects that the Cu-Cl-Cu exchange is ferromagnetic if the bond angle is close to 90$^o$, and antiferromagnetic if it deviates from $90^o$. However, when a Cu-Cl-Cu exchange path contains a long Cu--Cl bond (represented by Cl-Cu--Cl in Table II), the exchange becomes ferromagnetic even if the Cl-Cu--Cl angle deviates considerably from $90^o$ because the magnetic orbital is not contained in the long Cu--Cl bond. \cite{whangbo2} The Cu-Cl--Cl-Cu spin exchange should become more strongly antiferromagnetic with increasing Cu-Cl--Cl angle and shortening Cl--Cl contact distance so that the overlap between the Cl $3p$ orbitals in the Cl--Cl contact becomes large.\cite{whangbo2} The six exchanges considered in our study are listed in Table II and described in Fig. 4 (a). To evaluate these exchanges, we determine the relative energies of seven possible ordered spin states of (CuCl)LaNb$_2$O$_7$ (see Fig. 4 (b)) on the basis of density functional calculations employing the frozen-core projector augmented wave method \cite{blochl, kresse1} encoded in the VASP \cite{kresse2} with the generalized-gradient approximation (GGA) for the exchange-correlation functional \cite{perdew}. The GGA plus on-site repulsion U (GGA+U) method \cite{dudarev} with effective U = 4 eV was used to properly describe the strong electron correlation of the Cu 3d states. By mapping the relative energies of the seven states determined from the GGA+U calculations onto the corresponding energies determined from the spin Hamiltonian defined in terms of the six exchanges, we obtain their values listed in Table II. The fourth nearest neighbor (NN) interaction, $J_4$, of the Cu-Cl--Cl-Cu exchange type is the strongest and antiferromagnetic. The other fourth NN coupling $J_4^{'}$ is also antiferromagnetic, but is much weaker than $J_4$: $J_4^{'}/J_4$ = 0.18. This is because $J_4$ has a larger Cu-Cl--Cl angle and a shorter Cl--Cl distance than $J_4^{'}$: 164.9$^o$ and 3.835 \AA~for $J_4$ vs. 156.0$^o$ and 4.231 \AA~for $J_4^{'}$. All other $J$s are ferromagnetic. The strengths of the six exchanges decrease in the order,  $J_4 > J_{1a} > J_{2a} > J_4^{'} > J_{2b} > J_{1b}$. %Note that interactions via short Cu-Cl bonds are much stronger than those with long Cu--Cl bonds, as expected.

The fact that $J_4$ is both the strongest interaction and is antiferromagnetc explains the long standing mystery of the spin singlet formation in (CuCl)LaNb$_2$O$_7$; the distance between the Cu ions connected by $J_4$ is indeed $r =$ 8.533 \AA, consistent with the intradimer distance obtained by the isolated spin dimer model. 
To improve the model we considered the exchange paths $J_{1a}$ and $J_{2a}$ in addition to $J_{4}$ and fitted the inelastic neutron scattering data to the first moment sum rule for powder data \cite{Hohenberg1974} 
\begin{align}
\left\langle E(Q) \right\rangle & =\hbar^2\int_{\Omega}\int_{\omega} \omega S(\mathbf{Q},\omega)d\omega d\Omega \nonumber \\
%\end{align}
%\begin{align}
& \propto-\sum_{s}J_{s}\left\langle \mathbf{S}_{0}\cdot \mathbf{S}_{d_{s}}\right\rangle\left|f_{\mathrm{Cu}^{2+}}(Q)\right|^{2}\left(1-\frac{sin(Qd_{s})}{Qd_{s}}\right). \nonumber
\end{align}
In this equation the integration is over solid angle and energy, $J_{s}$ is the exchange constant coupling the $s^{th}$ nearest neighbor spins, $d_{s}$ is their separation and $\left\langle \textbf{S}_{0}.\textbf{S}_{d_{s}}\right\rangle$ is the two spin correlation function for this pair. From this expression it is clear that each coupling constant $J_{s}$ produces a modulation in the first moment with a periodicity depending on the separation $d_{s}$ of the spins. Therefore the dominant exchange interactions can be deduced but the magnitude of the exchange constant cannot be determined because the fitted quantity is the product of the exchange constant and the spin correlation function. The best fit was obtained with $J_4\left\langle \textbf{S}_{0}.\textbf{S}_{d_{4}}\right\rangle=0.027(2)$, $J_1a\left\langle \textbf{S}_{0}.\textbf{S}_{d_{1a}}\right\rangle=0.011(2)$ and $J_2a\left\langle \textbf{S}_{0}.\textbf{S}_{d_{2a}}\right\rangle=0.005(2)$. Note that the ground state obtained by the first principle calculations is the AF3 state shown in Fig. 4 (b), and thus a positive value of $J_s\left\langle \textbf{S}_{0}.\textbf{S}_{d_{s}}\right\rangle$ indicates that the spin correlation agrees with the sign of the exchange constant, i.e., the spins are parallel for a ferromagnetic interaction and antiparallel for an antiferromagnetic interaction. %as would be expected for the strongest exchange interactions.
The fit is much better than that of the simple dimer model, and the fitted parameters are consistent with our first principle calculation results. This indicates that in (CuCl)LaNb$_2$O$_7$ the fourth nearest neighbor Cu$^{2+}$ ions form spin singlets, which are arranged orthogonally in the $ab$-plane, while the coupling between them is primarily ferromagnetic. Thus, the spin lattice of (CuCl)LaNb$_2$O$_7$ is best described as ferromagnetically coupled Shastry-Sutherland quantum spin singlets. 
  
%\bibliography{reference}
% kageyama

\end{document}